# Bone Density and Texture from Minimally Post-Processed Knee Radiographs in Subjects with Knee Osteoarthritis

Jukka Hirvasniemi[1], Jaakko Niinimäki[2,3], Jérôme Thevenot[3], Simo Saarakkala[2,3]

[1]Center for Machine Vision and Signal Analysis, Faculty of Information Technology and Electrical Engineering, University of Oulu, Oulu, Finland
[2]Department of Diagnostic Radiology, Oulu University Hospital, Oulu, Finland
[3]Research Unit of Medical Imaging, Physics and Technology, Faculty of Medicine, University of Oulu, Oulu, Finland
`j.hirvasniemi@erasmusmc.nl`

**Abstract.** Plain radiography is the most common modality to assess the stage of osteoarthritis. Our aims were to assess the relationship of radiography-based bone density and texture between radiographs with minimal and clinical post-processing, and to compare the differences in bone characteristics between controls and subjects with knee osteoarthritis or medial tibial bone marrow lesions (BMLs). Tibial bone density and texture was evaluated from radiographs with both minimal and clinical post-processing in 109 subjects with and without osteoarthritis. Bone texture was evaluated using fractal signature analysis. Significant correlations ($p<0.001$) were found in all regions (between 0.94 and 0.97) for calibrated bone density between radiographs with minimal and clinical post-processing. Correlations varied between 0.51 and 0.97 ($p<0.001$) for $FD_{Ver}$ texture variable and between -0.10 and 0.97 for $FD_{Hor}$. Bone density and texture were different ($p<0.05$) between controls and subjects with osteoarthritis or BMLs mainly in medial tibial regions. When classifying healthy and osteoarthritic subjects using a machine learning-based elastic net model with bone characteristics, area under the receiver operating characteristics (ROCAUC) curve was 0.77. For classifying controls and subjects with BMLs, ROCAUC was 0.85. In conclusion, differences in bone density and texture can be assessed from knee radiographs when using minimal post-processing.

## 1 Introduction

Osteoarthritis (OA) is the most common degenerative joint disease and it causes a large economic burden to the society as the direct and indirect costs can reach as high as 2.5% of the gross domestic product of a nation [20], not to mention the reduction of the quality of life of an individual. OA-related changes in the subchondral bone include bone sclerosis (hardening of bone), osteophytes, bone cysts, and bone deformation [4].

Plain radiography is a cheap, fast, and widely available imaging method. It is especially suitable for imaging of bone tissue. Plain radiographs are commonly used in diagnostics of diseases that affect bone density and structure, such as OA. Due to the aforementioned advantages of the plain radiography, development of image analysis tools for the assessment of OA-related changes is of interest. However, efforts are needed to produce comparable plain radiographs between X-ray imaging systems from different manufacturers, as image acquisition settings and post-processing (PP) algorithms affect the appearance of the final image and, *e.g.*, the assessment of bone density [14]. Typical clinical PP algorithms apply non-linear filtering and adjustment on contrast curves of an image to improve diagnostic readability [14]. To overcome the issue with quantitative image analyses, calibration of the grayscale values in an image using an aluminum step wedge has been proposed [8, 14, 21, 32].

We have recently shown that bone texture assessed from radiographs differs between subjects with and without bone marrow lesions (BMLs)6. However, that study did not assess bone density due to the lack of a calibration object in images and the bone texture was calculated only from two regions of interests (ROIs) in medial tibia. Recently, multiple ROIs covering the majority of the proximal tibia area were proposed to address this limitation [10, 11].

In theory, texture analysis of bone is not as dependent on the imaging conditions as the direct evaluation of grayscale values. In OA research, fractal analysis is the most common method for the assessment of bone structure from plain radiographs [3, 6, 10, 11, 15, 16, 18, 19, 24]. To date, bone texture or density has not been assessed from clinical X-ray images with minimal PP and compared between controls and OA subjects. We believe that simultaneous assessment of bone density and structure from a plain radiograph would be an advantage. Furthermore, the results would be more comparable if the effect of PP algorithms is minimized, *i.e.*, by calculating the bone density and texture from X-ray images with minimal possible PP strength.

Consequently, the first aim of this study was to investigate the relationship of radiography-based bone density and texture between X-ray images with minimal PP and with default clinical PP algorithm to find out how much the PP algorithm affect these measurements. The second aim was to compare the differences in bone characteristics (density and texture) between controls and subjects with knee OA or medial tibial BMLs to find out whether the changes in bone characteristics can be detected from X-ray image with minimal PP. Finally, a machine learning model was built to assess how well subjects with and without OA or medial tibial BMLs can be discriminated based on their bone density and texture only.

## 2  Subjects and Methods

### 2.1  Study subjects

This cross-sectional study included 109 subjects (66 women, 43 men) with and without OA (Table 1). Written informed consent was obtained from each participant. The study was carried out in accordance with the Declaration of Helsinki and approved by the Ethical Committee of Northern Ostrobothnia Hospital District, Oulu University Hospital (number 7/2016).

**Table 1.** Description of the subjects ($n = 109$).

| Variable | Mean (Standard deviation) | Min – max |
|---|---|---|
| Anthropometric variables | | |
| Age (years) | 58.1 (6.0) | 45 – 68 |
| Height (m) | 1.70 (0.09) | 1.50 – 1.92 |
| Weight (kg) | 78.3 (14.2) | 50.0 – 127.6 |
| Body mass index (kg/m$^2$) | 27.2 (4.4) | 19.7 – 40.3 |
| KL grade distribution | | |
| KL 0 | 14 | |
| KL 1 | 43 | |
| KL 2 | 28 | |
| KL 3 | 22 | |
| KL 4 | 2 | |

### 2.2 Acquisition and grading of the radiographs

Bilateral posterior-anterior weight-bearing radiographs with knees in semi-flexion were acquired (DigitalDiagnost, Philips Medical Systems, 10 degrees X-ray beam angle, 60 kVp, automatic exposure, pixel size: 0.148 mm x 0.148 mm, source – detector distance: 153 cm) and processed with minimal PP and default clinical PP algorithm. Right knees of the subjects were used in the analyses. Three radiographs with minimal PP and two radiographs with default clinical PP were missing and, thus, the total numbers of radiographs with minimal and clinical PP were 106 and 107, respectively.

An experienced musculoskeletal radiologist (JN) classified the knees according to the KL grading [12], in which grade zero corresponds to a healthy knee and grade four to severe OA.

### 2.3 Selection of regions of interests

To assess bone density and texture from the radiographs, 18 ROIs were semi-automatically placed across the proximal tibia (Figure 1). The locations were identical in radiographs with minimal and default PP. Two ROIs (size: 14 mm x 6 mm) were placed into the subchondral bone in the middle of the medial and lateral tibial plateaus immediately below the cartilage – bone interface. Anatomical landmarks for the ROIs were tibial spine, subchondral bone plate, the dense subchondral trabecular bone, and outer borders of the proximal tibia. The locations and sizes of the ROIs were based on the previous literature [6, 7, 9-11]. A custom-made MATLAB software (version R2017b, The MathWorks, Inc., Natick, MA, USA) was used for the placement (JH) of the ROIs. We have previously shown that the reproducibility of the texture variables from the tibial subchondral and trabecular bone is high [7, 9].

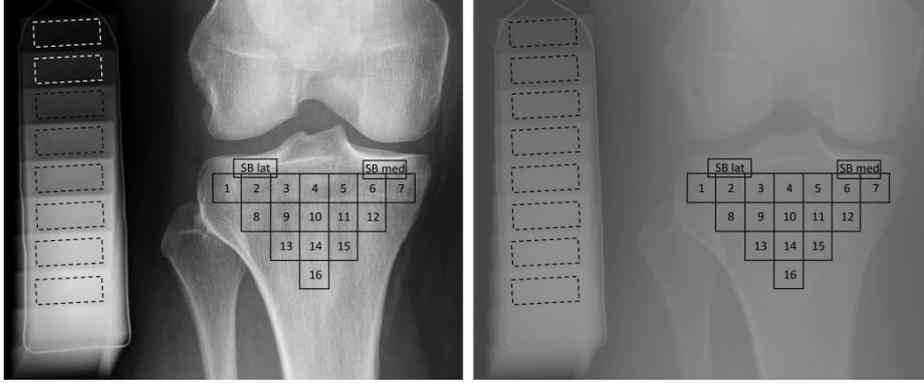

**Fig. 1**. Location of regions of interest (ROIs). The ROIs were exactly in the same location in images with default clinical post-processing (left) and with minimal post-processing (right). Two ROIs were placed in subchondral trabecular bone immediately below the cartilage-bone interface in the middle part of the medial and lateral tibial plateaus. Sixteen ROIs were placed under the dense subchondral trabecular bone area. Dashed rectangles show the areas where the mean value of the steps of the aluminum step wedge were calculated.

### 2.4 Bone density assessment

Two different methods to evaluate bone density were used, *i.e.*, 1) the mean grayscale value of the ROI (= GV) and 2) the aluminum step wedge thickness that corresponds to the measured GV (= $GV_{mmAl}$). The corresponding step wedge thickness was calculated by fitting a third order polynomial to the mean grayscale values of the eight first steps in the step wedge in each image and comparing the values of that fitted curve to the GV. The two thickest steps were omitted because the grayscale values were saturated at those steps. The step wedge was present in all images. For one subject, mean GV in medial subchondral bone ROI was higher than the highest grayscale value in step wedge and that ROI was therefore excluded from the analyses (extrapolation of the step wedge values would have been needed).

### 2.5 Bone texture

Fractal signature analysis (FSA) method was used to estimate fractal dimension [18, 19]. In brief, the image was dilated and eroded in horizontal and vertical directions with a rod-shaped one-pixel wide structuring element. After that, the volume, V, between dilated and eroded images was calculated. Calculations were repeated by varying the element length r from 2 to 7 pixels. The surface area, A(r), was obtained from the Equation 1:

$$A(r) = (V(r)-V(r-1))/2 \tag{1}$$

Subsequently, a log-log plot was constructed by plotting log of A(r) against log of r. Finally, the fractal dimension was estimated by fitting a regression line to points in the plot and local fractal dimensions were obtained at 0.30 mm, 0.44 mm, 0.59 mm, and 0.74 mm sizes. When the structuring element is pointing in the horizontal direction, fractal dimension of vertical structures ($FD_{Ver}$) is produced and vice versa. High fractal dimension values are associated with high complexity of the image, whereas low complexity results in low fractal dimension values.

### 2.6 Magnetic resonance imaging

Right knees of all but one subjects ($n = 108$) were scanned with a 3-Tesla magnetic resonance imaging (MRI) scanner (Siemens Skyra, Siemens Healthcare) using sagittal T2-weighted dual-echo steady-state (repetition time (TR): 14.1 ms, echo time (TE): 5 ms, echo train length (ETL): 2, pixel size: 0.6 mm x 0.6 mm, slice thickness: 0.6 mm), 3-D sagittal proton-density (PD)-weighted SPACE fat-suppressed turbo spin-echo (TSE) (TR: 1200 ms, TE: 26 ms, ETL: 49, pixel size: 0.6 mm x 0.6 mm, slice thickness: 0.6 mm), coronal PD-weighted TSE (TR: 2800 ms, TE: 33 ms, ETL: 4, pixel size: 0.4 mm x 0.4 mm, slice thickness: 3 mm), and coronal T1-weighted TSE (TR: 650 ms, TE: 18 ms, ETL: 2, pixel size: 0.4 mm x 0.4 mm, slice thickness: 3 mm) sequences. An experienced musculoskeletal radiologist (JN) assessed the presence of BMLs and a subject was included in the medial tibial BML group if he/she had any BML (including ill-defined lesions, bone marrow edema and subchondral cysts) in the medial anterior, central, or posterior part of tibia.

### 2.7 Statistical analyses

The normality of the variables was assessed using Shapiro-Wilk test. The relationship between normally distributed variables was evaluated using Pearson's correlation analysis ($r$) while Spearman's rank correlation ($r_s$) was applied if at least one of the variables was not normally distributed. Absolute values of correlation coefficients were interpreted as follows: 0.00 - 0.19 very weak, 0.20 - 0.39 weak, 0.40 - 0.59 moderate, 0.60 - 0.79 strong and 0.80 - 1.00 very strong correlation28. No adjustments for multiple comparisons were performed [26].

For comparing differences between controls (group 0), subjects with radiographic knee OA without medial tibial BML (group 1), and subjects with medial tibial BML (group 2), based on the normality of the variables either analysis of variance (ANOVA) or Kruskal-Wallis test was applied. These analyses were combined with post-hoc tests without correction for the Type I error rate across the pairwise tests and using Bonferroni correction. Clinical covariates were age, gender, and body mass index. Bone characteristics from X-rays images with minimal PP was used.

Machine learning was used for dimensionality reduction and to assess how well subjects with and without OA or BMLs can be discriminated based on their bone density and texture only. For this, a regularized logistic regression method called elastic net was used [5, 33]. The elastic net linearly combines the L1 and L2 penalties of lasso and ridge regression methods. To optimize the ratio of the L1 and L2 penalties ($\alpha$) and the strength of the penalty parameter ($\lambda$) of the elastic net, leave-one-out cross-validation was performed. When α is close to zero, the elastic net approaches ridge regression, while when α is 1, lasso regression is performed. The performance of the bone density and texture (from X-ray images with minimal PP) feature model to discriminate subjects with and without OA as well as subjects with and without medial tibial BMLs was assessed using area under the receiver operating characteristics curve (ROC AUC). Statistical analyses and elastic net experiments were done using R (version 3.1.2) software with Caret [17] (version 6.0), pROC [25] (version 1.8), glmnet [5] (version 2.0), and dunn.test (version 1.3.2) packages.

## 3 Results

### 3.1 Comparison of bone density and texture between minimal and default clinical PP

Without normalization of grayscale values in the reference step wedge, the correlations between GVs from X-ray images with minimal PP and default clinical PP varied from 0.18 ($p$=0.07) to 0.63 ($p$<0.001) depending on the ROI (Figure 2, Table 2). For the GV$mmAl$ variable, statistically significant (p<0.001) very strong correlations were found in all ROIs (between 0.94 and 0.97) (Figure 2, Table 2).

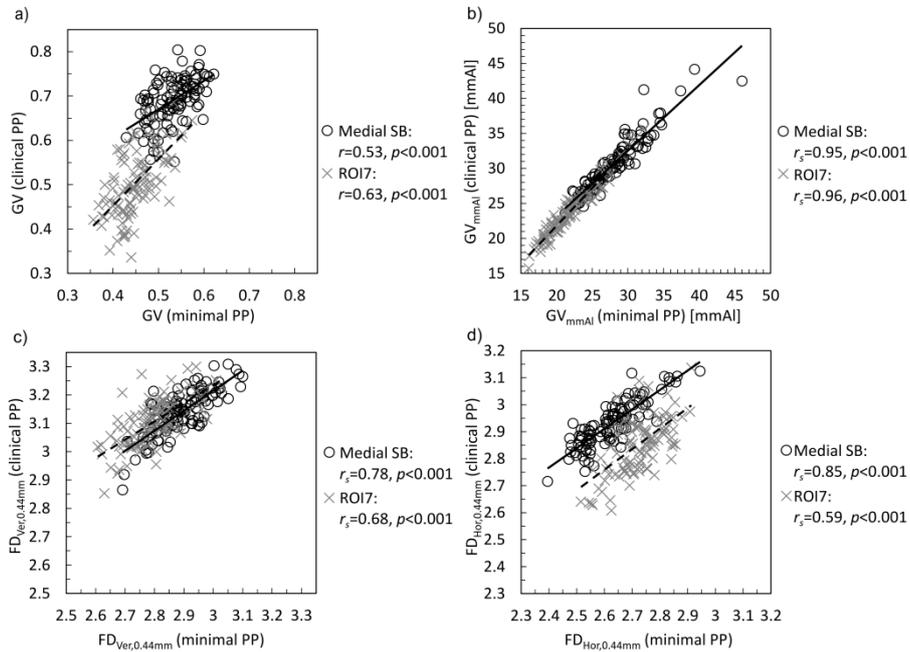

**Fig. 2.** Correlations between (a) GV, (b) GV$_{mmAl}$, (c) FD$_{Ver,0.44mm}$, and (d) FD$_{Hor,0.44mm}$ measured from X-ray images with minimal and default clinical post-processing (PP) in medial subchondral bone (SB) and ROI7. The scale varies between figures but is constant within a figure.

Moderate to very strong correlations (between 0.51 and 0.97, $p<0.001$) were found between X-ray images with minimal PP and default clinical PP when using $FD_{Ver}$ texture variable, whereas when using $FD_{Hor}$, the correlations varied at different scales and ROIs from no correlation to very strong correlation (between -0.10 and 0.97) (Figure 2, Table 2).

### 3.2    Differences in bone characteristics between controls, OA subjects, and subjects with medial tibial BMLs

Subjects with medial tibial BMLs (group 2) had significantly ($p<0.05$) higher body mass index than subjects with OA but without BMLs (group 1) or controls (Table 3). Moreover, subjects with medial tibial BMLs were older ($p<0.05$) than controls.

$GV_{mmAl}$ from X-ray images with minimal PP was significantly ($p<0.05$) higher in group 1 (OA without medial tibial BML) and in group 2 (medial tibial BML) than in control group in all medial side ROIs (subchondral bone ROI and ROI6, ROI7, and ROI12) (Table 3).

Statistically significant differences ($p<0.05$) in $FD_{Ver}$ (in all scales) from X-ray images with minimal PP in medial side ROIs were found. For example, $FD_{Ver,0.44mm}$ in subchondral bone and in ROI7 was significantly different among controls than in group 1 (OA without medial tibial BML) or group 2 (medial tibial BML) (Table 3). Statistically significant differences ($p<0.05$) in FDHor were found in medial and lateral side ROIs (Table 3).

### 3.3    Classification of OA or BML subjects and controls

A ROC AUC value of 0.77 (95% confidence interval (CI): 0.68 – 0.87) was obtained for classifying healthy and OA subjects using the elastic model with parameters describing bone density and texture from X-ray image with minimal PP (Figure 3a). The values for α and λ hyperparameters of the elastic model were 1 and 0.118, respectively. The bone density and texture parameters that were selected in the final model are shown in Table 4. A ROC AUC value of 0.81 (95% CI: 0.72 – 0.89) was obtained when covariates (age, gender, and body mass index) were included in the model (Figure 3a).

A ROC AUC value of 0.85 (95% CI: 0.76 – 0.95) was obtained for classifying controls and subjects with medial tibial BML using the elastic model with parameters describing bone density and texture (Figure 3b). The values for α and λ hyperparameters of the elastic model were 0.8 and 0.037, respectively. The bone density and texture parameters that were selected in the final model are shown in Table 5. A similar ROC AUC value of 0.85 (95% CI: 0.76 – 0.94) was obtained when covariates were included in the model (Figure 3b).

**Table 3.** Mean (standard deviation) values of the selected variables among controls, subjects with radiographic OA but no medial tibial BMLs, and subjects with medial tibial BMLs. Bone density and texture parameters were measured from X-ray images with minimal post-processing.

| Variable | Group 0: Controls (n = 52) | Group 1: OA, no medial tibial BML (n = 30) | Group 2: Medial tibial BML (n = 23) | $p$-value |
|---|---|---|---|---|
| Age (years) | 56.4 (6.3)[2] | 58.3 (5.5) | 60.8 (4.4) | 0.019[a] |
| Body mass index (kg/m$^2$) | 25.0 (2.5)[1,2] | 28.1 (3.8)[2] | 30.9 (5.8) | <0.001 |
| $GV_{mmAl}$ in medial SB (mmAl) | 26.9 (3.1)[1,2] | 29.2 (4.7) | 29.6 (4.5) | 0.011[a] |
| $GV_{mmAl}$ in ROI6 (mmAl) | 25.0 (2.6)[1,2] | 26.8 (3.4) | 26.8 (3.9) | 0.016 |
| $GV_{mmAl}$ in ROI7 (mmAl) | 20.1 (2.1)[1,2] | 22.3 (2.9)[2] | 24.0 (4.0) | <0.001 |
| $GV_{mmAl}$ in ROI12 (mmAl) | 23.6 (2.4)[1,2] | 25.4 (2.9) | 25.6 (3.7) | 0.006 |
| $FD_{Ver,0.30mm}$ in medial SB | 2.65 (0.09)[2] | 2.68 (0.08) | 2.71 (0.07) | <0.001 |
| $FD_{Ver,0.30mm}$ in ROI6 | 2.65 (0.08)[2] | 2.66 (0.07) | 2.70 (0.06) | 0.028 |
| $FD_{Ver,0.30mm}$ in ROI7 | 2.55 (0.06)[2] | 2.57 (0.06)[2] | 2.62 (0.06) | <0.001[a] |
| $FD_{Ver,0.30mm}$ in ROI12 | 2.70 (0.06)[2] | 2.71 (0.06) | 2.74 (0.06) | 0.043 |
| $FD_{Ver,0.30mm}$ in ROI15 | 2.70 (0.06)[2] | 2.72 (0.08)[2] | 2.74 (0.05) | 0.013[a] |
| $FD_{Ver,0.44mm}$ in medial SB | 2.86 (0.09)[1,2] | 2.90 (0.08)[2] | 2.95 (0.08) | <0.001 |
| $FD_{Ver,0.44mm}$ in ROI7 | 2.76 (0.08)[1,2] | 2.80 (0.09)[2] | 2.85 (0.08) | <0.001[a] |
| $FD_{Ver,0.59mm}$ in medial SB | 2.90 (0.11)[1,2] | 2.96 (0.12) | 3.01 (0.11) | <0.001 |
| $FD_{Ver,0.59mm}$ in ROI7 | 2.79 (0.11)[1,2] | 2.85 (0.12)[2] | 2.92 (0.12) | <0.001 |
| $FD_{Ver,0.59mm}$ in ROI12 | 3.14 (0.10)[1,2] | 3.19 (0.12) | 3.20 (0.07) | 0.024 |
| $FD_{Ver,0.74mm}$ in medial SB | 2.84 (0.11)[2] | 2.90 (0.16)[2] | 2.96 (0.13) | 0.002[a] |
| $FD_{Ver,0.74mm}$ in ROI7 | 2.70 (0.15)[1,2] | 2.77 (0.15) | 2.85 (0.17) | <0.001 |
| $FD_{Ver,0.74mm}$ in ROI12 | 3.08 (0.13)[1,2] | 3.16 (0.13) | 3.17 (0.13) | 0.005 |
| $FD_{Hor,0.30mm}$ in lateral SB | 2.52 (0.08)[2] | 2.55 (0.10) | 2.57 (0.08) | 0.046 |
| $FD_{Hor,0.30mm}$ in ROI7 | 2.57 (0.08)[2] | 2.59 (0.08) | 2.62 (0.07) | 0.025 |
| $FD_{Hor,0.59mm}$ in ROI2 | 2.96 (0.07) | 2.98 (0.05)[2] | 2.93 (0.10) | 0.020 |
| $FD_{Hor,0.59mm}$ in ROI3 | 2.98 (0.07)[2] | 2.97 (0.07)[2] | 2.91 (0.09) | <0.001[a] |
| $FD_{Hor,0.74mm}$ in ROI1 | 2.92 (0.06)[2] | 2.92 (0.08)[2] | 2.87 (0.08) | 0.013 |
| $FD_{Hor,0.74mm}$ in ROI2 | 2.97 (0.08)[2] | 2.98 (0.14)[2] | 2.91 (0.14) | 0.040 |
| $FD_{Hor,0.74mm}$ in ROI3 | 2.98 (0.07)[2] | 2.97 (0.11)[2] | 2.91 (0.11) | 0.004[a] |
| $FD_{Hor,0.74mm}$ in ROI7 | 2.77 (0.08)[2] | 2.75 (0.11)[2] | 2.70 (0.11) | 0.012[a] |

SB = subchondral bone, ROI = region of interest, $GV_{mmAl}$ = mean grayscale value calibrated with aluminum step wedge, FD = fractal dimension of vertical (Ver) or horizontal (Hor) structures, [a] = differences tested using Kruskal-Wallis test. Numbers in superscript means significant differences between groups without correction of $p$-values. Bolded numbers means significant differences between groups using Bonferroni post-hoc test.

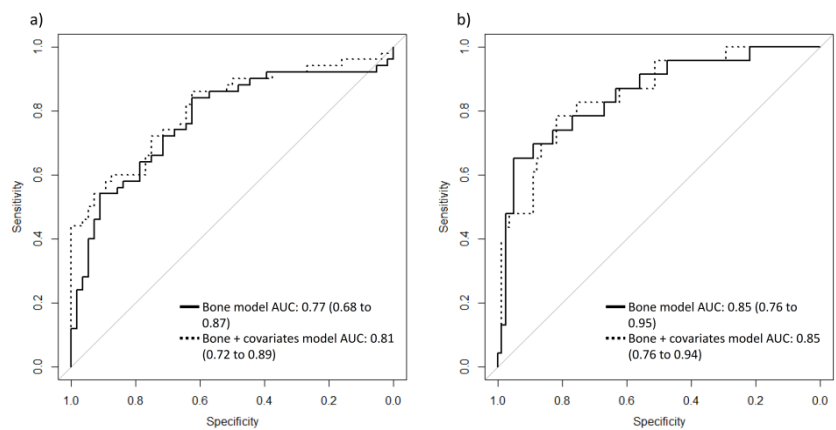

**Fig. 3**. Receiver operating characteristics curves and respective area under the curve (AUC) values for discriminating a) subjects without and with radiographic knee osteoarthritis as well as b) subjects without and with medial tibial bone marrow lesions using models that included bone characteristics (bone density and texture) from X-ray images with minimal post-processing and bone characteristics combined with covariates (age, gender, body mass index).

**Table 4**. Bone density and texture parameters from X-ray images with minimal post-processing in the elastic net model to discriminate healthy ($n = 56$) and subjects with radiographic knee osteoarthritis ($n = 50$).

| Variable | Coefficient |
| --- | --- |
| Intercept | -0.111 |
| $GV_{mmAl}$ in ROI7 | 0.470 |
| $FD_{Ver,0.59mm}$ in medial SB | 0.003 |
| $FD_{Ver,0.44mm}$ in ROI7 | 0.174 |

SB = subchondral bone, ROI = region of interest, GVmmAl = mean grayscale value calibrated with aluminum step wedge, FDVer = fractal dimension of vertical structures.

**Table 5**. Bone parameters from X-ray images with minimal post-processing in the elastic net model to discriminate subjects without ($n = 82$) and with medial tibial bone marrow lesion ($n = 23$).

| Variable | Coefficient |
| --- | --- |
| Intercept | -1.752 |
| $GV_{mmAl}$ in ROI7 | 0.220 |
| $FD_{Ver,0.44mm}$ in medial SB | 0.396 |
| $FD_{Ver,0.74mm}$ in medial SB | 0.004 |
| $FD_{Ver,0.30mm}$ in ROI7 | 0.260 |
| $FD_{Ver,0.30mm}$ in ROI12 | 0.127 |
| $FD_{Ver,0.30mm}$ in ROI15 | 0.393 |
| $FD_{Ver,0.59mm}$ in ROI7 | 0.197 |
| $FD_{Ver,0.74mm}$ in ROI4 | -0.092 |
| $FD_{Ver,0.74mm}$ in ROI6 | -0.126 |
| $FD_{Hor,0.59mm}$ in ROI2 | -0.012 |
| $FD_{Hor,0.59mm}$ in ROI3 | -0.644 |
| $FD_{Hor,0.59mm}$ in ROI13 | -0.091 |
| $FD_{Hor,0.74mm}$ in ROI1 | -0.351 |
| $FD_{Hor,0.74mm}$ in ROI2 | -0.012 |
| $FD_{Hor,0.74mm}$ in ROI5 | -0.313 |
| $FD_{Hor,0.74mm}$ in ROI7 | -0.213 |
| $FD_{Hor,0.74mm}$ in ROI8 | -0.443 |
| $FD_{Hor,0.74mm}$ in ROI12 | -0.243 |

SB = subchondral bone, ROI = region of interest, $GV_{mmAl}$ = mean grayscale value calibrated with aluminum step wedge, FD = fractal dimension of vertical (Ver) or horizontal (Hor) structures.

## 4 Discussion

This study evaluated bone density and texture from knee X-ray images with minimal PP. First, the association of bone density and texture between X-ray images with minimal PP and default clinical PP was assessed. Our results show that bone density was strongly correlated between these two PP methods when the grayscale values were calibrated with the reference step wedge. Correlations of bone texture parameters varied from weak to very strong. Second, we assessed bone density and bone texture from X-ray images with minimal PP, and significant differences between controls (group 0), subjects with OA but without medial tibial BMLs (group 1), and subjects with medial tibial BMLs (group 2) were found. Third, machine learning based elastic net model showed that both bone density and texture parameters contributed to the model when discriminating controls and subjects with OA or subjects with BMLs. Furthermore, relatively good ROC AUC values to discriminate subjects without and with OA (0.77), as well as without and with BMLs (0.85), using bone density and texture parameters were obtained.

Strong correlations were obtained when the grayscale values were calibrated, whereas the correlations between the grayscale values without calibration were weak or moderate. Based on this and earlier results, calibration of grayscale values are required when assessing bone density from plain radiographs [8, 14]. Varying correlations in texture parameters between X-ray images were found. One reason for this may be that the clinical PP algorithm applies non-linear filtering and adjusts contrast curves of an image and, *e.g.*, edges in the image are enhanced. The appearance of the bone contours and trabeculae is different between these two images. The lower correlation were found especially in $FD_{Ver}$ and $FD_{Hor}$ parameters at larger scales (0.59 mm or 0.74 mm) and may due to different appearance of the bone trabeculae. Our results indicate that when assessing bone texture at larger scales, the effect of PP should be considered.

Differences in bone density and texture between controls and subjects with OA without medial tibial BMLs as well as subjects with medial tibial BMLs were found. Bone density was higher among subjects with OA and among subjects with BMLs than among controls in medial side ROIs. Bone sclerosis may be one reason for the higher bone density values. Differences in bone texture between groups using $FD_{Ver}$ was observed in medial side ROIs while $FD_{Hor}$ was significantly different in some lateral side ROIs also. This result show that the bone structure was different between groups. In our earlier study, we showed that, *e.g.*, $FD_{Ver}$ was associated with 3-dimensional connection and separation of the bone trabeculae [8]. The finding that bone density and texture differs between controls and OA subjects is in line with previous studies using plain knee radiographs with clinical PP algorithm [3, 7, 18, 21-23]. The finding for the bone density, however, contradicts for one study in which no association between KL grade and radiography-based bone density in knee was found [13]. Our previous study revealed that bone texture assessed from radiographs differs between subjects with and without bone marrow lesions (BMLs), but bone density was not assessed in that study [6]. In general, our present results demonstrate that bone density and texture can be assessed from X-ray images with minimal PP to detect differences between controls, subjects with OA, and subjects with BMLs.

To our knowledge, this is the first study that assessed bone density and texture from X-ray images with minimal PP among subjects with OA or BMLs. Because the direct evaluation of grayscale values of a radiograph is problematic, calibration of the grayscale values using an aluminum step wedge has been proposed [8, 14, 21, 32]. In an earlier study, bone density in human cadaver tibia was assessed from X-ray image with minimal PP and a strong correlation to actual bone mineral density assessed with dual X-ray absorptiometry was reported [14]. Another study with human cadaver tibias showed that radiography-based tibial bone density and texture are related with the actual 3-dimensional structure and amount of bone [8].

Elastic net models were used to assess how well subjects with and without OA or BMLs can be discriminated based on their bone density and texture. Leave-one-out cross-validation was used in order to find optimal hyperparameters for the models. The elastic net also reduces the dimensionality of the feature vector, which was necessary because initially all bone density and texture parameters from all ROIs were fed into the model. The ROC AUC values to discriminate subjects without and with OA as well as without and with medial tibial BMLs using bone density and texture parameters were relatively high and when covariates were included in the model, the classification performance was slightly improved in discriminating subjects without and with OA, but not for discriminating subjects without and with BMLs. The results are in line with previous studies, although they used plain knee radiographs with clinical PP algorithm [29, 31]. One study reported an accuracy of 85.4% for discriminating healthy and OA subjects using bone texture from plain knee radiographs. They used signature dissimilarity method to obtain bone texture. Another study reported a ROC AUC of 0.74 for discriminating healthy and OA subjects using directional fractal signature method [29].

It should be noted that a perfect classification was not expected in this study. This is because bone texture does not actually directly affect the KL grading, yet marginal osteophytes, bone sclerosis, cysts, deformation of bone, and narrowing of the joint space are considered in it. Furthermore, BMLs were assessed from MRI data. Thus, it can be that some subjects with OA do not actually have changes in their subchondral or trabecular bone. The use of KL grade as ground truth was justified, because it is the gold standard when clinically assessing the level of OA. When aiming to automatically assess the KL grade, the entire joint area should be fed in the model [1, 2, 27, 30]. However, in this study we wanted to specifically evaluate the changes in bone density and texture.

This study has some limitations. First, bone density and texture parameters are quantitative and continuous, whereas KL grading and BML evaluation are semi-quantitative, subjective, and discrete. Furthermore, bone texture is not directly evaluated in KL grading. Second, our data was cross-sectional and, thus, we were unable to assess how well bone density and texture predict the development or progression of OA. Third, in future studies higher number of X-rays with minimal PP is desired. It should be noted, however, that to our knowledge this is the first study that acquired X-rays with minimal PP using human subjects without and with OA or BMLs.

In conclusion, PP algorithm did have effect on the grayscale values and texture parameters, especially on fractal dimensions with larger scales. Differences in bone density and texture, assessed from X-ray images with minimal PP, were found between controls, subjects with OA but without BMLs, and subjects with medial tibial BMLs. Finally, relatively good classification between controls and OA subjects as well as controls and subjects with medial tibial BML using only bone density and texture parameters was obtained. Our results indicate that calibration of grayscale values are required when assessing bone density from plain radiographs and the effect of PP should be considered when assessing bone texture at larger scales.

## Acknowledgments

The research leading to these results has received funding from the Academy of Finland (projects 268378 and 308165), European Research Council under the European Union's Seventh Framework Programme (FP/2007-2013) / ERC Grant Agreement no. 336267, and Business Finland: Finnish Funding Agency for Innovation (grant number 1241/31/2016). The funding sources had no role in the study design, data collection or analysis, interpretation of data, writing of the manuscript, or in the decision to submit the manuscript for publication.## References

1. Antony, J., K. McGuinness, K. Moran, and N. E. O'Connor. Automatic detection of knee joints and quantification of knee osteoarthritis severity using convolutional neural networks. Lect. Notes Comput. Sci. 10358 LNAI:376-390, 2017.
2. Antony, J., K. McGuinness, N. E. O'Connor, and K. Moran. Quantifying radiographic knee osteoarthritis severity using deep convolutional neural networks. Proc. Int. Conf. Pattern Recognit. 1195-1200, 2017.
3. Buckland-Wright, C. Subchondral bone changes in hand and knee osteoarthritis detected by radiography. Osteoarthritis Cartilage 12:S10-9, 2004.
4. Buckwalter, J. A. and H. J. Mankin. Articular cartilage: degeneration and osteoarthritis, repair, regeneration, and transplantation. Instr. Course Lect. 47:487-504, 1998.
5. Friedman, J., T. Hastie, and R. Tibshirani. Regularization Paths for Generalized Linear Models via Coordinate Descent. J. Stat. Softw 33:1-22, 2010.
6. Hirvasniemi, J., J. Thevenot, A. Guermazi, J. Podlipská, F. W. Roemer, M. T. Nieminen, and S. Saarakkala. Differences in tibial subchondral bone structure evaluated using plain radiographs between knees with and without cartilage damage or bone marrow lesions-the Oulu Knee Osteoarthritis study. Eur. Radiol. 27:4874-4882, 2017.
7. Hirvasniemi, J., J. Thevenot, V. Immonen, T. Liikavainio, P. Pulkkinen, T. Jämsä, J. Arokoski, and S. Saarakkala. Quantification of differences in bone texture from plain radiographs in knees with and without osteoarthritis. Osteoarthritis Cartilage 22:1724-1731, 2014.
8. Hirvasniemi, J., J. Thevenot, H. T. Kokkonen, M. A. Finnilä, M. S. Venäläinen, T. Jämsä, R. K. Korhonen, J. Töyräs, and S. Saarakkala. Correlation of Subchondral Bone Density and Structure from Plain Radiographs with Micro Computed Tomography Ex Vivo. Ann. Biomed. Eng. 44:1698-1709, 2016.
9. Hirvasniemi, J., J. Thevenot, J. Multanen, M. Haapea, A. Heinonen, M. T. Nieminen, and S. Saarakkala. Association between radiography-based subchondral bone structure and MRI-based cartilage composition in postmenopausal women with mild osteoarthritis. Osteoarthritis Cartilage 25:2039-2046, 2017.
10. Janvier, T., R. Jennane, H. Toumi, and E. Lespessailles. Subchondral tibial bone texture predicts the incidence of radiographic knee osteoarthritis: data from the Osteoarthritis Initiative. Osteoarthritis Cartilage 25:2047-2054, 2017.
11. Janvier, T., R. Jennane, A. Valery, K. Harrar, M. Delplanque, C. Lelong, D. Loeuille, H. Toumi, and E. Lespessailles. Subchondral tibial bone texture analysis predicts knee osteoarthritis progression: data from the Osteoarthritis Initiative. Osteoarthritis Cartilage 25:259-266, 2017.
12. Kellgren, J. H. and J. S. Lawrence. Radiological assessment of osteo-arthrosis. Ann. Rheum. Dis. 16:494-502, 1957.
13. Kinds, M., K. Vincken, E. Vignon, S. T. Wolde, J. Bijlsma, P. Welsing, A. Marijnissen, and F. Lafeber. Radiographic features of knee and hip osteoarthritis represent characteristics of an individual, in addition to severity of osteoarthritis. Scand. J. Rheumatol. 41:141-149, 2012.

**Table 2.** Correlations between bone characteristics (density and texture parameters) measured from X-ray images with minimal and default clinical post-processing. $n = 103 – 104$.

| ROI | GV | $GV_{mmAl}$ | $FD_{Ver,0.30mm}$ | $FD_{Ver,0.44mm}$ | $FD_{Ver,0.59mm}$ | $FD_{Ver,0.74mm}$ | $FD_{Hor,0.30mm}$ | $FD_{Hor,0.44mm}$ | $FD_{Hor,0.59mm}$ | $FD_{Hor,0.74mm}$ |
|---|---|---|---|---|---|---|---|---|---|---|
| SB medial  | 0.53** | 0.95** | 0.81** | 0.78** | 0.78** | 0.66** | 0.91** | 0.85** | 0.88** | -0.10 |
| SB lateral | 0.39** | 0.94** | 0.94** | 0.87** | 0.81** | 0.85** | 0.95** | 0.89** | 0.90** | 0.25** |
| ROI1  | 0.23*  | 0.94** | 0.94** | 0.87** | 0.77** | 0.70** | 0.96** | 0.62** | 0.35** | 0.24* |
| ROI2  | 0.36** | 0.96** | 0.96** | 0.97** | 0.93** | 0.90** | 0.97** | 0.68** | 0.52** | 0.48** |
| ROI3  | 0.46** | 0.96** | 0.96** | 0.94** | 0.91** | 0.90** | 0.82** | 0.56** | 0.25* | 0.13 |
| ROI4  | 0.47** | 0.97** | 0.96** | 0.97** | 0.96** | 0.94** | 0.72** | 0.64** | 0.43** | 0.35** |
| ROI5  | 0.42** | 0.97** | 0.94** | 0.96** | 0.95** | 0.93** | 0.73** | 0.66** | 0.47** | 0.45** |
| ROI6  | 0.41** | 0.96** | 0.97** | 0.95** | 0.93** | 0.90** | 0.76** | 0.66** | 0.49** | 0.35** |
| ROI7  | 0.63** | 0.96** | 0.88** | 0.68** | 0.51** | 0.64** | 0.96** | 0.59** | 0.29** | 0.13 |
| ROI8  | 0.32** | 0.96** | 0.96** | 0.97** | 0.95** | 0.94** | 0.76** | 0.61** | 0.33** | 0.23* |
| ROI9  | 0.36** | 0.97** | 0.96** | 0.95** | 0.95** | 0.95** | 0.70** | 0.70** | 0.48** | 0.40** |
| ROI10 | 0.35** | 0.97** | 0.94** | 0.97** | 0.97** | 0.94** | 0.67** | 0.69** | 0.51** | 0.42** |
| ROI11 | 0.29** | 0.97** | 0.96** | 0.93** | 0.89** | 0.86** | 0.65** | 0.72** | 0.50** | 0.42** |
| ROI12 | 0.33** | 0.96** | 0.93** | 0.86** | 0.70** | 0.62** | 0.60** | 0.74** | 0.46** | 0.30** |
| ROI13 | 0.39** | 0.97** | 0.95** | 0.94** | 0.96** | 0.95** | 0.69** | 0.75** | 0.61** | 0.49** |
| ROI14 | 0.33** | 0.96** | 0.95** | 0.97** | 0.95** | 0.95** | 0.69** | 0.66** | 0.55** | 0.47** |
| ROI15 | 0.18   | 0.97** | 0.92** | 0.94** | 0.90** | 0.82** | 0.77** | 0.71** | 0.49** | 0.48** |
| ROI16 | 0.27** | 0.95** | 0.95** | 0.96** | 0.96** | 0.95** | 0.60** | 0.68** | 0.46** | 0.29** |

\*\*$p < 0.001$, \*$p < 0.05$, ROI = region of interest, SB = subchondral bone, GV = mean grayscale value of the ROI, $GV_{mmAl}$ = GV calibrated with aluminum step wedge, FD = fractal dimension of vertical (Ver) or horizontal (Hor) structures.